\documentclass[twocolumn,showpacs,aps,pre]{revtex4}
\usepackage{graphicx}
\usepackage{url}
[1999/03/29 PSNFSS v.7.2
Times font as default roman
: S Rahtz]

\begin{document}
\renewcommand{\thefootnote}{\fnsymbol{footnote}}
\sloppy
\newcommand{\rp}{\right)}
\newcommand{\lp}{\left(}
\newcommand \be  {\begin{equation}}
\newcommand \ba {\begin{eqnarray}}
\newcommand \ee  {\end{equation}}
\newcommand \ea {\end{eqnarray}}

\title{Predicting Failure using Conditioning on Damage History:
\\ Demonstration on Percolation and Hierarchical Fiber Bundles} 
\thispagestyle{empty}

\author{J.V. Andersen$^{1,2}$ and D. Sornette$^{2,3}$}
\affiliation{$^1$ U.F.R. de Sciences Economiques, Gestion, Math\'ematiques 
et Informatique, CNRS UMR 7536 and Universit\'e Paris X-Nanterre, 92001
Nanterre Cedex}
\affiliation{$^2$ Laboratoire de Physique de la Mati\`ere Condens\'ee,
CNRS UMR 6622 and Universit\'e de Nice-Sophia Antipolis, 06108
Nice Cedex 2, France}
\affiliation{$^3$ Institute of Geophysics and Planetary Physics
and Department of Earth and Space Sciences,
University of California, Los Angeles, CA 90095}

\email{vitting@unice.fr, sornette@moho.ess.ucla.edu}

\date{\today}

\begin{abstract}
We formulate the problem of probabilistic predictions of global failure
in the simplest possible model based on site percolation
and on one of the simplest model
of time-dependent rupture, a hierarchical fiber bundle model. 
We show that conditioning the predictions on the knowledge of the
current degree of damage (occupancy density $p$ or number and size
of cracks) and on some
information on the largest cluster improves significantly the
prediction accuracy, in particular by allowing to identify those
realizations which have anomalously low or large clusters (cracks).
We quantify the prediction gains using two measures, the
relative specific information gain (which is the variation of 
entropy obtained by adding new information) and the root-mean-square
of the prediction errors over a large ensemble of realizations.
The bulk of our simulations have been obtained 
with the two-dimensional site percolation model
on a lattice of size $L \times L=20 \times 20$ and hold true for other
lattice sizes. For the hierarchical fiber bundle model,
conditioning the measures of damage on the
information of the location and size of the largest crack extends
significantly the critical region and the prediction skills.
These examples illustrate how on-going damage can be used
as a revelation of both the realization-dependent pre-existing
heterogeneity and the damage scenario undertaken by each
specific sample.

\end{abstract}


\maketitle

\section{Introduction}

The idea underlying this paper was inspired by the method of ``reverse
tracing of precursors'' (RTP) introduced in Refs.~\cite{KB1,KB2} as a
method of earthquake prediction based on seismicity patterns. In a
nutshell, the RTP method consists first in delineating a spatial domain
${\cal S}(t)$ by using a space-time correlation analysis of past seismicity
up to the present time $t$ and then in constructing precursory diagnostics based on
past seismicity restricted to this spatial domain ${\cal S}(t)$ (called chains 
in Refs.~\cite{KB1,KB2}). In Refs.~\cite{KB1,KB2}, the precursory functions
used to issue a prediction are based on previously documented seismic anomalies
(see \cite{KBsolo} and references therein) and will not be our concern. 
Rather, the question we are asking is what could justify the innovation
presented in Refs.~\cite{KB1,KB2} to constrain the construction of 
precursory diagnostics to some special spatial domains recognized from
some spatio-temporal correlation analysis of past seismicity? Indeed, 
Refs.~\cite{KB1,KB2} do not provide an explanation on
why their method should work and what could be its underlying 
physical mechanism(s), since their approach is based on the
pragmatic mathematical pattern recognition method initiated long ago by
Gelfand et al. \cite{Gelfand}. Our paper is the first one in a series
which shows how the idea behind the RTP can be actually justified on
physical grounds and used for improving previous prediction methods for
earthquakes or material ruptures.

We first present the problem and explore its implications for the
percolation model and then test the robustness of the results 
and extend them to a time-dependent hierarchical fiber bundle model. 

\section{Formulation of the percolation model}

As a first step, we propose to formulate the problem with perhaps the
simplest model of heterogenous media undergoing a transition, the site
percolation model \cite{Deutscher,Grimmett,Stauffer}. By doing so,
we aim at capturing the essence of the idea. 

Consider a
two-dimensional lattice of $L \times L$ sites which are initially empty.
We then fill one by one the sites at random positions and denote by $p$
the corresponding fraction of occupied sites. Any given realization
${\cal C}_L$ will be characterized by some threshold $p_c({\cal C}_L)$
at which the occupied sites form a cluster which barely percolates from
one side of the system to its opposite side. It is known that, for $L
\to \infty$, $p_c({\cal C}_L)$ becomes independent of the specific
realization of the system and converge to a unique number $p_c^{\infty}
= 0.5927460 \pm 0.0000005$ \cite{Ziff}. It is also well-known
that, for finite $L$, $p_c({\cal C}_L)$ is a random number distributed according to 
a probability density function (PDF) $P(p_c)$ 
centered on a value shifted downwards from $p_c^{\infty}$ by an amount
and with a width which are both proportional to 
$1/L^{1/\nu}$, where $\nu=4/3$ is the universal exponent (in two-dimensions)
of the correlation length, defined roughly speaking as the typical size of the largest
cluster. The shift and width of the PDF $P(p_c)$ are characteristic of the
so-called finite-size scaling of the critical percolation transition \cite{Hu1}.

For our purpose which is to relate with the prediction of a rupture or an
earthquake, we interpret $p$ as the running time, which is also the
fraction of the lattice which is damaged. We thus envision 
the two-dimensional lattice as being progressively damaged at a rate
of one site failing per unit time. The percolation threshold
$p_c({\cal C}_L)$ then corresponds to the time when there 
is a connected path of damaged sites running from one side to the other,
such that the system is deconnected
into at least two pieces, a diagnostic of rupture. Hence, the progressive
filling of the sites in the percolation problem described above
corresponds to the progressive damage of an initially pristine system.

Roux et al. \cite{rouxetal88} have shown that rupture is equivalent to
percolation in the limit of very large disorder and, by extension,
rupture processes can be considered as nothing but (complicated)
correlated percolation problems \cite{criticalSorbook,reviewYip}. Since,
by definition, the addition of new sites in percolation model has no
interaction, correlation or memory of the past, the formulation of the
idea inspired by the RTP method in this context necessarily reduces the
scope of the approach. This is because the information present in real rupture and
earthquake cases based on correlation and memory in the time domain has
no bearing in the prediction of the percolation threshold $p_c({\cal C}_L)$.
In subsequent papers, we will investigate different
examples of ``correlated'' percolation, namely models of rupture, 
in which time-dependent precursors can be coupled with the 
spatial organization of damage.

\section{Predictions of the percolation threshold}

\subsection{A hierarchy of prediction levels \label{hieraleves}}

Suppose that a given realization in a system of size $L \times L$ is at
the cumulative fraction $p$ of damaged sites. What level of prediction
is possible for its percolation threshold $p_c({\cal C}_L)$?
We now describe different levels of prediction of the
percolation threshold based on increasing the available information.
\begin{enumerate}
\item The first level of prediction is what we call the
unconditional prediction, which amounts to not even use the knowledge
that the system has the cumulative fraction $p$ of damaged sites.
It corresponds to the statistical distribution of $p_c({\cal C}_L)$.
This is the information available at the beginning of a simulation.

\item The second level of prediction is to use the fact that we want to
predict $p_c({\cal C}_L)$ conditioned on the fact that we know that the
system has reached the cumulative fraction $p$ of damaged sites.
It is obvious that this improves on the first level: for instance, if
by luck, $p$ happens to be already quite large (say larger than the
average of $p_c({\cal C}_L)$) and the system is still
not percolating, then we know for sure that the value of $p_c({\cal C}_L)$
for this system will be larger than $p$.

\item The third level of prediction incorporates additional 
information on how the damage over the $p L^2$ sites is organized.
For instance, typical experiments of rupture have access to the
spatial organization of acoustic emissions, which provide clues
on the localization of damage. In this spirit, suppose that 
we can measure the fraction of damaged sites belonging to the
larger cluster at $p$ or the size along the horizontal and 
vertical directions of the larger cluster. Then, this should give
us some additional information to improve on the prediction.
Indeed, if we measure for two given realizations that the larger
cluster has a horizontal size close to $L$ in the first
one and $L/2$ in the second one for a given $p$, 
we can guess that the first system will in general percolate
sooner (for a smaller $p_c({\cal C}_L)$) than the second system.

\item One can imagine many other levels of prediction using
all kinds of additional information, such as the statistics of 
the clusters, their shape, positions, etc.

\item The last ultimate level of prediction is to use
all the information on the exact locations of all damaged sites
and condition the prediction of $p_c({\cal C}_L)$ on this knowledge.
\end{enumerate}

In the following, we implement the first three levels of predictions
and show that we obtain substantial gains at the third level. This is
perhaps not surprising, but this provides a quantitative demonstration
on how prediction can be improved by using information on the spatial
organization of damage. Additionally, it tells us what are the limits
of predictability, given each level of information.

\subsection{First and second prediction levels}

\begin{figure}
\includegraphics[width=14cm]{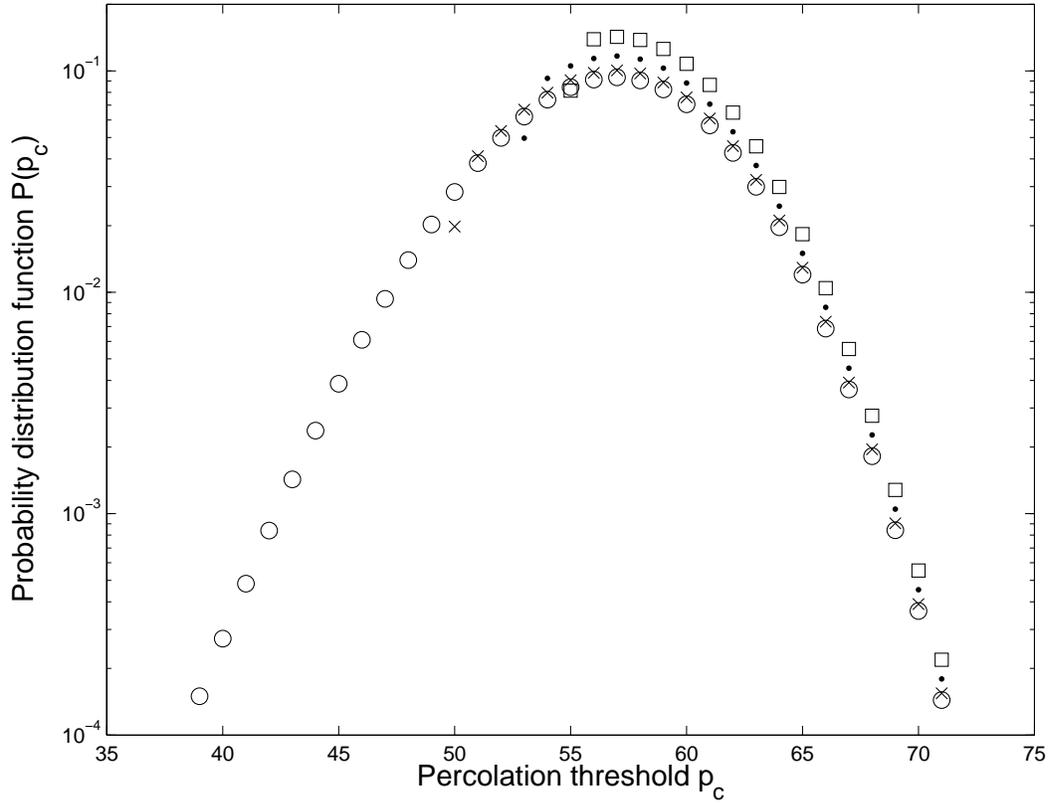}
\caption{\protect\label{Fig1} Circles:
standard probability distribution function (PDF)
$P_L(p_c)$ as a function of the percolation threshold $p_c$ (in percent) for $L=20$. 
Crosses: conditional PDF $P_L(p_c|p=0.5)$ conditioned on those systems 
which have not percolated for a fixed occupation density $p=0.5$.
Dots: conditional PDF $P_L(p_c|p=0.53)$ conditioned on those systems 
which have not percolated for a fixed occupation density $p=0.53$.
Squares: conditional PDF $P_L(p_c|p=0.55)$ conditioned on those systems 
which have not percolated for a fixed occupation density $p=0.55$.
}
\end{figure}

The first prediction level described in section \ref{hieraleves} amounts
to constructing the standard probability distribution function (PDF)
$P_L(p_c)$ of the percolation thresholds, shown by the circles in Figure
\ref{Fig1} for $L=20$. We have used $50$ million realizations to get
a good statistics. Such distribution is the standard tool
for the study of finite-size scaling \cite{Hu1}. For our purpose,
it quantifies the range of predictions for the percolation thresholds
$p_c({\cal C}_L)$ in the form of a probabilistic forecast.

Crosses, dots and squares 
show the second prediction level, corresponding to the
PDF's $P_L(p_c|p)$ conditioned on those systems 
which have not percolated for a fixed occupation density $p=0.50, 0.53$ and 
$p=0.55$ respectively. Since for $L=20$, the unconditional PDF $P_L(p_c)$
is quite broad with as many as $40 \%$ of the realizations
percolating with $p_c({\cal C}_L) < 0.55$, the condition that 
$p_c({\cal C}_L)$ has to be larger than $0.55$ transforms $P_L(p_c)$ into 
a significantly more peaked conditional PDF $P_L(p_c|p=0.55)$.
In the language of the prediction problem, the PDF's $P_L(p_c|p)$
shown with the crosses, dots and squares provide the probabilistic forecasts
for rupture, available at ``time'' $p$ and conditioned only on the knowledge of $p$.

\subsection{Third prediction level}

We implement the third prediction level described in in section \ref{hieraleves} 
in two ways. Let us call $p_{\xi}(p)$ the fraction of sites belonging
to the largest cluster and $\xi(p)$
the largest of the linear size projected on the $x$ and $y$ axes of the largest cluster
within the system when the occupancy density is $p$.

\begin{figure}
\includegraphics[width=14cm]{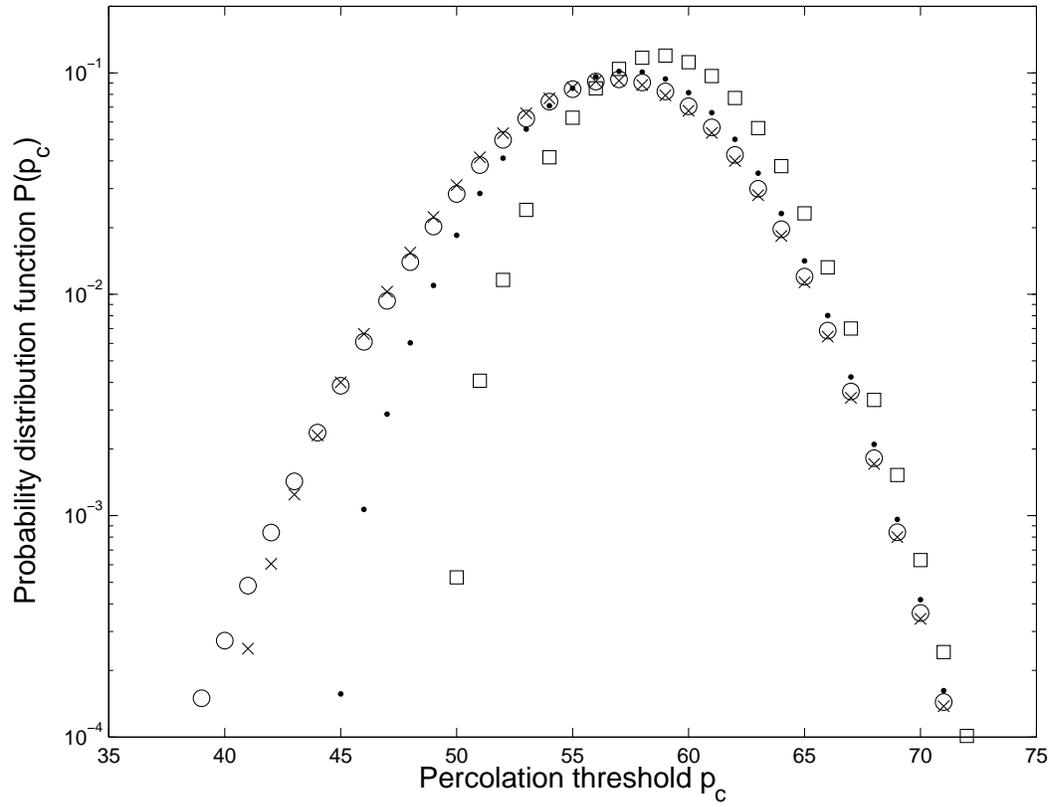}
\caption{\protect\label{Fig2} PDF $P_L(p_c|p, p_{\xi})$ 
as a function of $p_c$ (in percent)
conditioned on both $p$ and $p_{\xi}=6\%$, where
$p_{\xi}$ is the fraction of sites belonging to the largest cluster,
for different values of $p$ (crosses: $p=0.4$, dots: $p=0.45$, squares: $p=0.5$). 
For comparison, the unconditional distribution of the first prediction level
is also shown with open circles.
}
\end{figure}

Figure \ref{Fig2} presents the PDF $P_L(p_c|p, p_{\xi})$ 
conditioned on both $p$ and $p_{\xi=6\%}$, 
for different values of $p$ (crosses: $p=0.4$, dots: $p=0.45$, squares: $p=0.5$). 
For comparison, the unconditional distribution of the first prediction level
is also shown with open circles. The gradual shift of the PDF 
$P_L(p_c|p, p_{\xi})$ to larger 
values of $p_c$ for increasing $p$ shows that the measurement of 
the largest cluster size which is fixed at a given  $p$ 
in the percolating process makes it 
more likely to see percolation occuring at ``late times'' (i.e. for large 
$p_c$'s) the larger the value of $p$. Intuitively, this just means that 
if one observes in two different systems for different values of $p$ 
the {\em same} concentration of the  largest 
cluster, the system with the largest value of $p$ is more likely 
to percolate at a later time.  
The shift and narrowing of the PDF's are clear 
illustration of the information one can gain by conditioning on 
relevant variables. 

\begin{figure}
\includegraphics[width=14cm]{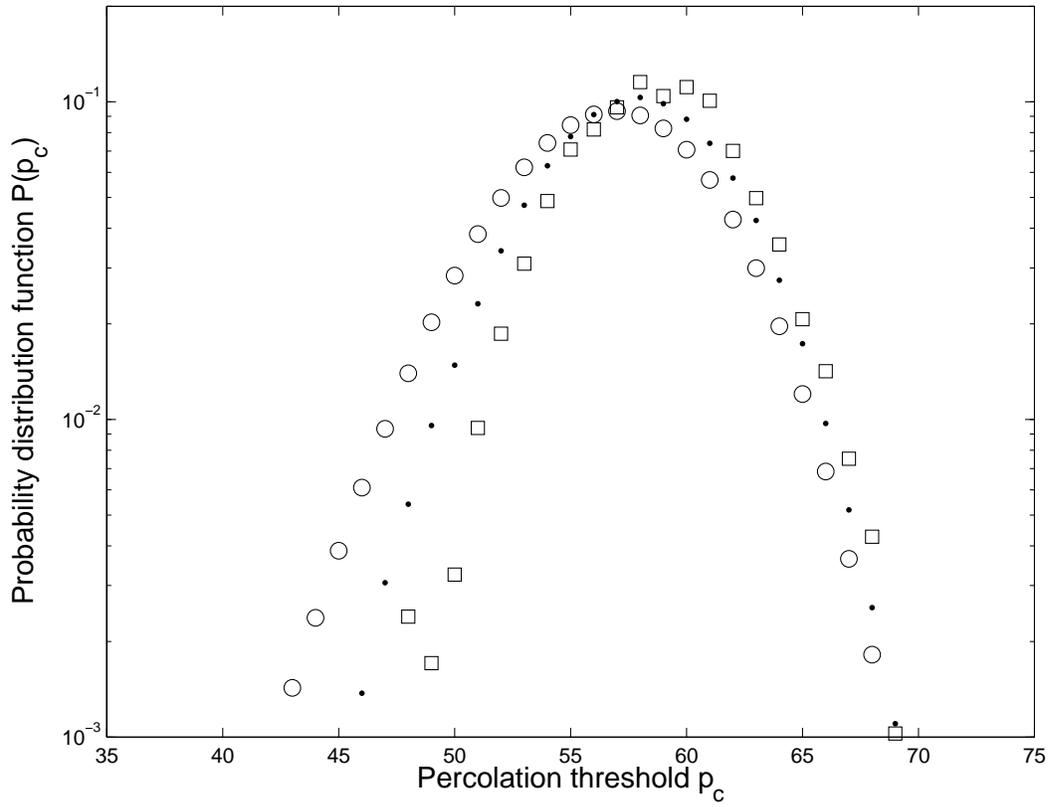}
\caption{\protect\label{Fig3} PDF $P_L(p_c|p, \xi)$ 
as a function of $p_c$ (in percent)
conditioned on both $p$ and $\xi=0.2 L$, where
$\xi$ is the largest of the linear size projected on 
the $x$ and $y$ axes of the largest cluster
within the system, for different
values of $p$ (dots: $p=0.35$, squares: $p=0.4$). 
For comparison, the unconditional distribution of the first prediction level
is also shown with open circles. 
}
\end{figure}

Figure \ref{Fig3} presents the PDF $P_L(p_c|p, \xi)$ 
conditioned on both $p$ and $\xi=0.2 L$ for different
values of $p$ (dots: $p=0.35$, squares: $p=0.4$). 
For comparison, the unconditional distribution of the first prediction level
is also shown with open circles. 
The results are similar to those presented in Figure \ref{Fig2}, with
a gradual shift and narrowing of the conditional PDF's to larger 
values of $p_c$ for increasing $p$. 
Using the largest projected cluster should give even more information on
the final value of $p_c$ for a given system, since e.g. two 
systems with the same $p$ and $p_{\xi}$, 
but one having a more elongated largest cluster than the other, 
should help the former reach the percolation threshold sooner on average.

Figure {Fig3b} shows as Figure \ref{Fig3} the PDF $P_L(p_c|p, \xi)$ 
for a fixed $p=40\%$ and different values of $\xi$:
$\xi/L=0.04$ (crosses),  $\xi/L=0.06$ (dots),
$\xi/L=0.08$ (squares) and $\xi/L=0.1$ (triangles). The open circles
represent the unconditional PDF $P_L(p_c)$ for comparison.

\section{Measures of goodness of the third level predictions}

\subsection{Information gain}

This standard measure of the improvement in the quality 
of forecasts when going from the first to the third prediction level is
the information gain $H-H_{sc}$, where 
$H$ is the unconditional entropy defined by
\be
  H  =   - \int P(p_c) \ln{(P(p_c))} d p_c~.
  \label{mghkes}
\ee
We consider two possible conditional entropies
$H_{sc}(p,p_{\xi})$ and $H_{sc}(p,\xi)$ 
associated with the two conditional schemes of the third level prediction
discussed in the previous section:
\be
 H_{sc}(p, p_{\xi} {\rm ~or~} \xi)  =  - 
 \int P(p_c|p,p_{\xi} {\rm~or~} \xi) \ln{(P(p_c|p,p_{\xi} {\rm~or~} \xi))}~.
\ee
The  relative ``specific information gain'' $I(p,p_{\xi} {\rm~or~} \xi)$ 
is then defined by
\be
I(p,p_{\xi} {\rm~or~} \xi)  \equiv {1 \over H}~ \left(H-H_{sc}(p,p_{\xi} {\rm~or~} \xi)\right)~.
\ee

Figure \ref{Fig4} shows $I(p,p_{\xi})$ (panel a)) and
$I(p,\xi)$ (panel b)) as a function of $p$ for 
various values of $p_{\xi}$ and $\xi$. The relative specific
information gains $I(p,p_{\xi})$ and $I(p,\xi)$ have qualitatively
the same behavior, characterized by three regimes.
\begin{enumerate}
\item For small values of $p$ (the smaller $p_{\xi}$ or $\xi$,
the smaller the values of $p$ for which this regime holds), we observe
some information gain when adding the information on $p_{\xi}$ or $\xi$.
This information gain can be ascribed to the realizations which
initially (i.e. for small $p$) have 
an abnormal large value of $p_{\xi}$ or $\xi$, and therefore are likely to 
percolate before the typical behavior. The knowledge of these
anomalously large $p_{\xi}$ or $\xi$, when they occur, gives an
improvement for the prediction of the percolation of these systems.
Translated in the context of the prediction of rupture, the information
gain shown in Figure \ref{Fig4} for small $p$'s is based on
the detection of anomalous cracks or defects at an early stage.
It is important to stress that the information gain is not uniform
over all realizations: most realizations are not much more predictable
by adding the information on  $p_{\xi}$ or $\xi$) for small $p$'s; only those
which have anomalous defects can be better predicted. This
result is reasonable and retrieves the standard approach in 
the applications of mechanical engineering to the prediction of rupture
in which the major efforts are put in the detection of possible
initial flaws in the material or structure.

\item For intermediate values of $p$, the 
information gain obtained by conditioning on $p_{\xi}$ or $\xi$
is limited if not negative, since for these values of $p$ the 
imposed $p_{\xi}$ or $\xi$ correspond to ``normal'' values.

\item Finally, for the larger $p$'s, the information gain accelerates
and become large since it becomes very unlikely to observe systems 
with such small values of $p_{\xi}$ or $\xi$. Therefore, the
knowledge that a given realization has an anomalously small $p_{\xi}$ or $\xi$
provides a highly meaningful information that percolation will
require a much large value of $p$ than the current value.
\end{enumerate}
While the relative specific
information gains $I(p,p_{\xi})$ and $I(p,\xi)$ have qualitatively
the same behavior, the gain is much larger for the later compared
with the former: this is because the geometrical size of the
larger cluster is much more relevant for percolation than 
the total number of sites in the large cluster.

\subsection{RMS of prediction errors}

We now quantify the errors of the prediction of the 
realization specific percolation threshold $p_c({\cal C}_L)$
based on the conditioning on $p$ and $p_{\xi} {\rm~or~} \xi$.
We imagine a situation mimicking a real life situation
in which one monitors the cumulative level of damage $p$
of a sample as well as the largest crack in the system.
Conditioned on the knowledge of $p$ and $p_{\xi} {\rm~or~} \xi$
for a given realization, how well can we predict the
rupture time $p_c({\cal C}_L)$ of the sample?

In order to address this question, we have first made $50$ million
realizations of system sizes $L=20$ to obtain a good estimate of the
conditional distributions $P(p_c|p,p_{\xi})$ and $P(p_c|p,\xi)$, 
which will be our prediction tools.
Having sampled these conditional distributions, we 
then constructed additional realizations that we monitored
to measure their $p_{\xi}(p)$ and $\xi(p)$ as a function of $p$.
For a given realization at a given $p$, knowing 
the corresponding specific $p_{\xi}(p)$, our prediction
is nothing but $P(p_c|p,p_{\xi}(p))$. Similarly, 
for a given realization at a given $p$, knowing 
the corresponding specific $\xi(p)$, our prediction
is nothing but $P(p_c|p,\xi(p))$. Note that our forecast
are intrinsically probabilistic, by construction. However, each
probabilistic forecast
can be translated into a single predicted number 
$p_c^{\rm predicted} (p)$,
for instance, the median of $P(p_c|p,p_{\xi}(p))$ or $P(p_c|p,\xi(p))$,
complemented with an uncertainty given by some measure of the 
width of these distributions (standard deviation or quantiles).

In order to assess the quality of such predictions, 
we need to construct statistics over ensembles of forecasts.
In addition, we would like to study how the quality of the 
predictions evolve with the degree of damage $p$, in particular
to test if we get advanced warning and how the prediction improves
or deteriorates as a function of $p$. Since, for each $p$,
we have two distributions of $p_{\xi} {\rm and} \xi)$ which move with $p$,
the amount of data to visualize is too large to remain comprehensible.
We propose to focus on fixed quantiles $q$ of the distributions of 
$p_{\xi} {\rm and} \xi)$, say $q=5\%$ and $q=95\%$, 
so that we issue predictions based on the pairs $p, p_{\xi}^{q}(p)$
(and similarly $p, \xi^{q}(p)$) where $p_{\xi}^{q}(p)$ 
(resp. $\xi^{q}(p)$) is the $q$-th 
quantile of the distribution of $p_{\xi}$ (resp. $\xi$) 
for the cumulative damage $p$.

For such a prediction, we can assess its error
by constructing the RMS (root-mean-square) of errors
\be
Q(p) \equiv  \langle  (p_c^{\rm predicted} (p)  -   p_c^{\rm true} )^2 \rangle^{1/2}~,
\label{nvyhwy}
\ee 
where $p_c^{\rm true}$ is the true value observed for the given 
system and where $p_c^{\rm predicted} (p)$ is our predicted 
value of $p_c$ for a given system and for a given $p$ 
and using a given quantile $q$ of the distribution of $p_{\xi}$ (resp. $\xi^{q}$) 
for the cumulative damage $p$. As our 
prediction $p_c^{\rm predicted} (p)$ 
for $p_c$, we have used the median value of the conditional 
cumulative distribution defined by
\be
P_{\le}(p_c^{\rm predicted}|p,p_{\xi}^{q}(p) {\rm~or~} \xi^{q}(p)) = 1/2~.
\label{mgkkle;sa}
\ee 

Figure \ref{Fig5b} shows
$Q(p, \xi)$ for $q=5\%$ and $q=95\%$, when using $P(p_c|p,\xi^q(p))$ as the predictor,
as a function of $p$.  The triangles correspond to $q=5\%$, the dots to
$q=95\%$, the crosses to $q=50\%$, while the circles show $Q(p)$ obtained 
using the conditioning only on $p$ for comparison. The correspond figure when using
$P(p_c|p,p_{\xi}(p))$ as the predictor is very similar and is thus not shown.
Figure \ref{Fig5ba} shows the gain in RMS
$Q(p) - Q(p, \xi)$ when adding the information on $\xi$.
These figures show the result of an implementation which 
mimics a real experiment of a material progressively brought to failure:
one would for a given time (that is $p)$ measure the
largest crack and, from the PDF's
documented from earlier experiments, get an estimate of $p_c$
corresponding to that $p, \xi$. Notice that all three estimates in figure \ref{Fig5b}
coincide for small $p$'s. The reason is of course that the PDF's for small
$p$ are very close to each other, whether the conditioning is
on $\xi$ (corresponding to a small
value of $p$) or just conditioned on $p$ itself.

These figures confirm the signicant gain in prediction accuracy
when conditioning the forecast on the $q=5\%$ and $q=95\%$ quantiles
of the distribution of $\xi$ (resp. $p_{\xi}$). The $q=5\%$ quantile
selects those realizations such that their largest cluster
is so small that $95\%$ of the realizations have a bigger largest cluster.
Conditioning on this information gives a significant gain in the
forecast, especially for advanced warnings. The improvement
deteriorates when $p$ approaches the average percolation threshold
and even changes sign with a worse quality for $p$ larger than about $54\%$.
We observe the opposite trend when conditioning on the 
$q=95\%$ quantile of the distribution of $\xi$, corresponding
to those realizations which have an anomalously big largest
cluster so that only $5\%$ of the realizations
have a bigger largest cluster. In this case, the prediction accuracy is improved
above for $p > 0.45$.

\section{Hierarchical Fiber Rupture Model with Time-Dependence}

The principles underlying the results on the percolation model presented
above are of general validity. Our following papers will investigate their
application and extension to other model systems and to different real
systems including concrete engineering systems (material failure,
structural collapse) and geophysical systems (earthquakes, landslides).
However, it is worthwhile already to present preliminary results obtained
on a more realistic (even still highly simplified) model
of damage evolution and rupture, to illustrate our point.

\subsection{Definition of the hierarchical bundle model}

The model describes the time evolution of damage leading to the
culminating global failure of a bundle of fibers in a creep experiment.
The model has been studied in \cite{Newmandyn,Newmandyn2,SSS}.
Consider a hierarchical bundle of elastic fibers subjected
to a constant stress load $\sigma$ per fiber applied at time $t=0$. 
The topology of the system is as follows. Each fiber
is associated with another fiber in a pair. Then, two neighboring pairs
are associated to each other, forming a pair of two pairs, and so on
iteratively up in a sequence of levels, thus defining a discrete hierarchical tree of local
coordination $2$. A system containing $n$ such levels has $2^n$ fibers. 
This topology impacts the dynamics of fiber rupture in the following way.
When one of the two fibers
of a given pair fails, its stress load is transfered instantaneously to the surviving
fiber, such that its load is doubled. When this fibers breaks, its
load is transfered to the pair of fibers associated to it if this second pair is still present.
Otherwise, it is transfered to the pair of two pairs linked at the next hierarchical
level. The last ingredient of the model is to specify how a fiber fails under
a given stress load history. Given some
stress history $s(t'), t'\geq0$, a fiber is assumed to break at some fixed random time,
where the probability that this random time takes a specific value $t$ is
specified by its cumulative distribution function 
\be
P_0(t)\equiv\int_0^t p_0(t')dt'=1-\exp\left\{-\kappa\int_0^t
[\sigma(t')]^{\rho} {\rm d}t'\right\}~.
\label{pdfggh}
\ee
This law captures the physics of stress corrosion and of failure 
due to stress-assisted thermal activation and progressive damage.
A system of $2^n$ fibers is fully specified by attributed to each
fiber $i=1, ..., 2^n$ at the beginning of the experiment a fixed failure
time $t_i$ taken from the distribution (\ref{pdfggh}). The failure time $t_i$
is by definition the time at which the fiber $i$ would have broken
if the stress had stayed constant equal to the initial value $\sigma$.
But, the fibers are coupled through the hierarchical load transfer rule 
defined above. As a consequence of the
hierarchical structure of the load transfers occuring at each rupture, the stress applied
to a given fiber may increase, leading to a shortening of its lifetime.

Let us consider quantitatively the effect of the rupture of one fiber
at time $t_1$ on the other fiber of its pair, which would have broken at time $t_2$ without
this additional load transfer. For a population of such pairs of fibers, the
distribution of the time-to-failure for the remaining fiber is obtained from
(\ref{pdfggh}) by taking the stress equal to $\sigma$ up to $t_1$ and
equal to $2\sigma$ from $t_1$ up to the second rupture, which now
occurs at a time $t_{12} < t_2$ itself function of $t_1$ and $t_2$:
\be
P_0(t_{12}) =1 - \exp\left\{-\kappa \sigma^{\rho} [t_1 + 2^{\rho} (t_{12}-t_1)]
\right\}~.
\label{pdfpoiggh}
\ee
Doing this calculation for the ensemble, the population of fibers must be
the same since the population is homogeneous at this level and $P_0(t_{12})$ should
therefore also be equal to $1 - \exp\left(-\kappa \sigma^{\rho} t_2 \right)$.
Considering that $t_{12}$ is a function of $t_2$, and
identifying this expression with (\ref{pdfpoiggh}), we re-derive the fundamental
result \cite{Newmandyn} that the time-to-failure of a fiber is modified from its initial value
$t_2$ to a smaller failure time $t_{12}$ by the influence of the other fiber which has
failed at the earlier time $t_1$, according to:
\be
t_{12}=t_1+2^{-\rho}(t_2-t_1)~ .
\label{fytuj}
\ee
The inequality $2^{-\rho} \leq 1$ (for $\rho > 0$) ensures that $t_1 \leq t_{12} \leq t_2$.
This corresponds to a genuine cooperative process as the time-of-failure of the
second fiber is decreased by the load transfer from the first fiber.
This remarkable result holds for any realization of
the stochastic process. Let us stress that this result now applies
not only at the level of individual fibers but at all levels within the hierarchy:
if $t_1$ and $t_2$ are the lifetimes of 
two uncoupled bundles, then (\ref{fytuj}) describes the effect of the rupture of the
first bundle on the second one which sees its load doubling at time
$t_1$. The relation (\ref{fytuj}) forms the basis for analytical as well as
numerical simulations. In particular, 
an exact Monte Carlo calculation of the probability distribution of
failure times of this hierarchical system indicates that the distribution of failure
times for the whole system is renormalized from $P_0(t)$ into
a staircase (or jumps from $0$ to $1$) at a well-defined non-zero
critical time $t^*$, as the system size $n$ tends to infinity, according
to a generalized central limit theorem.
It has also been shown theoretically and numerically that the
rate of fiber failures diverges (up
to finite size effects) according to a power law $\sim 1/(t^*-t)^{p(\rho)}$
upon the approach to the global rupture time $t^*$ for
$\rho > 1$, where $p$ depends on $\rho$ \cite{Newmandyn2,SSS}.
In our investigation below, we take $\kappa=1$ and $\rho=2$. 

\subsection{Third level prediction by conditioning the distribution of lifetimes
on the observation of the large crack}

We address the central question of this paper, namely, how
the revelation of information up to the present in the form of the 
partial knowledge of where and when fibers or groups of fibers have broken
may be exploited to bracket better and better the realization-specific
lifetime of a whole given system.

In order to mimic a real-life situation, we consider a creep experiment
of our hierarchical fiber system such that, at time $0$, a stress $\sigma$
is applied. We have no access to the specific individual lifetimes 
of the individual constituting fibers, only to their PDF $p_0(x)$.
At time passes, damage occurs, that is, fibers break,
thus revealing their initial lifetimes. The situation becomes
of course complicated because of the interactions between the fibers
through the hierarchical stress-load defining the model,
as the damage spreads accross the levels
of the hierarchy. In a real-life experiment, the damage would be
measured for instance by acoustic emissions, with both time and
space localization giving information
of which fibers have been broken and at what time.

Our goal here is to construct schemes that uses
some information in space and time on the damage that occured
until time $t$ to form a better prediction for the rupture
of the next level of the hierarchy and for the whole system,
in the form of a PDF of lifetimes for the total system.

Figure \ref{fighierspacetime} gives an illustration of the 
space-time evolution of fiber damage for a system of
$2^8=256$ fibers. One can observe a transition from
initial random uncorrelated ruptures to a progressive
organization with growth of ``cracks''and fusion between ``cracks''
associated with the acceleration of damage up to the culmination
global failure.

Now, suppose that we observe the evolution of such a system from
time $0$ to some ``present'' time $t$, before complete failure. 
Furthermore, suppose that our measurement is imperfect and 
we do not have access to all the information on the position 
and times of individual fiber failures. Let us assume that we only know
the size $2^m$ of the larger crack (or bundle) that has broken up to
time $t$ and some addition information on
the fibers that broke within this crack at earlier times. 
Is this knowledge useful? Figure \ref{cumcond} shows two
different measures of the cumulative number of broken fibers as a
function of $t$ (in log-log scales) for a given realization. The thick curve shows the
unconditional cumulative number of broken fibers. The thin curve shows,
as a function of time $t$, the cumulative number of broken fibers, which
broke either within the largest crack or within its complement in their
pair within the hierarchy. It is worth emphasizing that the time-evolution of both cumulative 
damage is knowable at each time $t$. One can observe a striking difference,
illustrating vividly the impact of conditioning on some available
partial information on the on-going damage, in order to improve
the prediction of the global failure: 
in the absence of conditioning (we count all broken fibers), one can observe
mostly a linear increase and, only at the very end, can one see an 
acceleration (which is a power law of $1/(t_c-t)$ as shown in the inset); 
In contrast, with the conditioning on the largest crack and its complement, the power law
regime is extended to very early time. 

This result can not be stressed sufficiently:
in the past two decades, material failure of heterogeneous materials have been
shown to belong to the class of dynamic critical phenomena (see
for instance the review \cite{reviewYip} and references therein), but the
critical region is in general difficult to observe and rather reduced 
in practical situation, thus hindering the applications (this is why other
techniques have been developed to enhance the predictability by extending
the region over which critical information can be extracted \cite{critrupjoh,sorpred}).
What is remarkable in Figure \ref{cumcond} is that, focusing on the largest
current crack and its neighborhood enhances the critical region tremendously,
thus offering a large potential for prediction at early times.

Figure \ref{Qtimehier} is the equivalent for the hierarchical rupture
model of figure \ref{Fig5b} previously constructed for the percolation model.
It shows the root-mean-square (rms) of the error or difference between
predictions of the global rupture time and the true realized one, 
for $5$ distinct prediction schemes using different conditioning.
The improvement due to conditioning is qualitatively similar but quantitatily
stronger than for the percolation model. This can be expected since
the hierarchical bundle model has a dynamics in which the failure times
of fibers keep the memory of past ruptures: the failure of a fiber
is a function of all the previous ruptures that impacted the load
history on this fiber. In constrast, the rupture of a bond in the
percolation is absolutely independent of past damage (except for the fact
that the rupture occurs on remaining intact bonds, which is the mechanism
underlying the benefits of conditioning exploited in previous sections).
The existence of memory is expected and one can verify that it improves
the prediction performance: we conjecture more generally that,
the larger the connectivity and interactions
between elements, the better should be the improvement of prediction
quality with conditioning upon new information.

\section{Concluding remarks}

Our goal has been to demonstrate that one can predict the 
percolation or rupture threshold, based on the knowledge of the amount of the current
damage and on some information on the largest cluster or crack in the
system. This problem was inspired by the idea of constructing
better predictors for earthquakes and ruptures based
on a combination of the space and time organization of damage.
In this paper, which is the first of a series, we have first considered
perhaps the worst and most difficult case for prediction, namely 
percolation, because in this model damage has no memory of the past
and not space-time correlation exist other than the 
properties associated with the geometry of connectivity.
Similar results, not shown here, have been obtained for other 
lattice sizes $L=10$ and $L=30, 40$ and $50$. 

Then, we have illustrated the robustness of the results
presented for the percolation model on one of the simplest model
of time-dependent rupture, a hierarchical fiber bundle model. 
We have shown that conditioning the measures of damage on the
information of the location and size of the largest crack extends
significantly the critical region and the prediction skills.

We will show in subsequent papers that the predictions obtained
in more realistic models of rupture which include realistic
correlation in the space-time organization of damage and of cracks
are significantly better, still. But our goal has been reached
here by showing that, in the worst possible and most difficult
case for prediction, we can achieve significant gains
by implementing the conditioning of some information on the spatial
organization of damage. In our practical implementation, we have considered 
the simplest information and many other algorithms can be developed
to improve on our results. This will be developed in future papers.

{\bf Acknowledgements}: We are grateful to V. Keilis-Borok
and I. Zaliapin for stimulating discussions.

{}

\clearpage

\clearpage

\clearpage


\clearpage

\begin{figure}
\includegraphics[width=14cm]{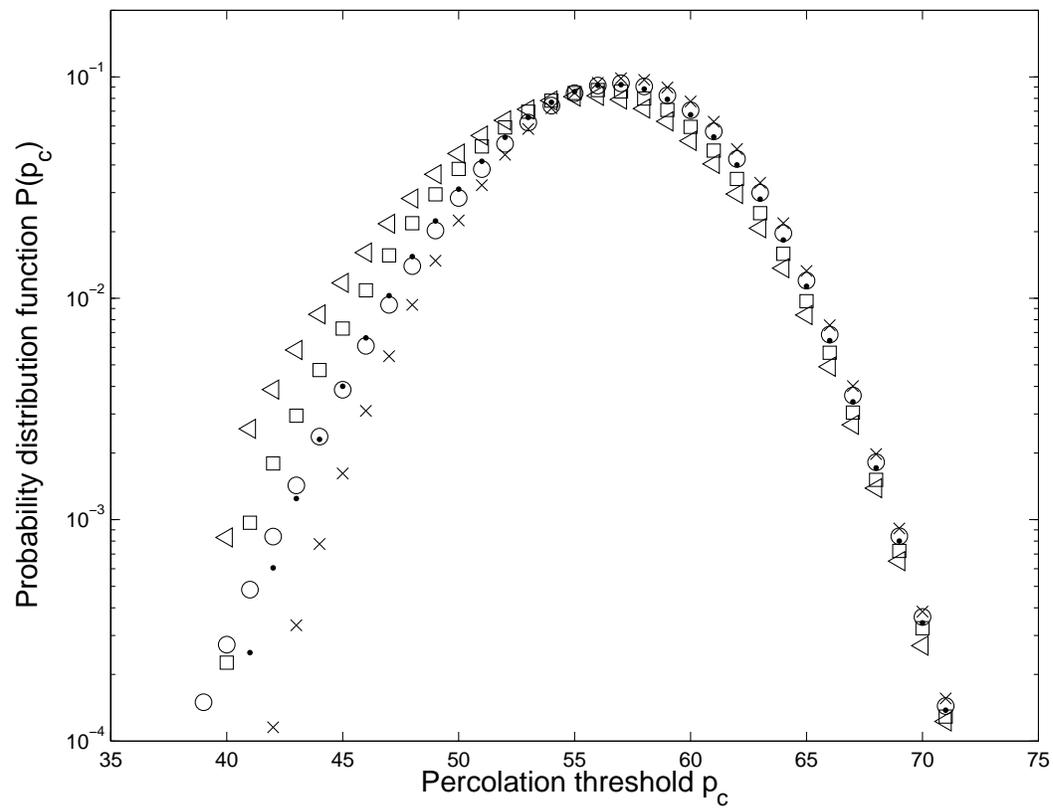}
\caption{\protect\label{Fig3b} Same as Figure \ref{Fig3}
for a fixed $p=40\%$ and different values of $\xi$:
$\xi/L=0.04$ (crosses),  $\xi/L=0.06$ (dots),
$\xi/L=0.08$ (squares) and $\xi/L=0.1$ (triangles). The open circles
represent the unconditional PDF $P_L(p_c)$ for reference.
}
\end{figure}

\clearpage

\begin{figure}
\includegraphics[width=14cm]{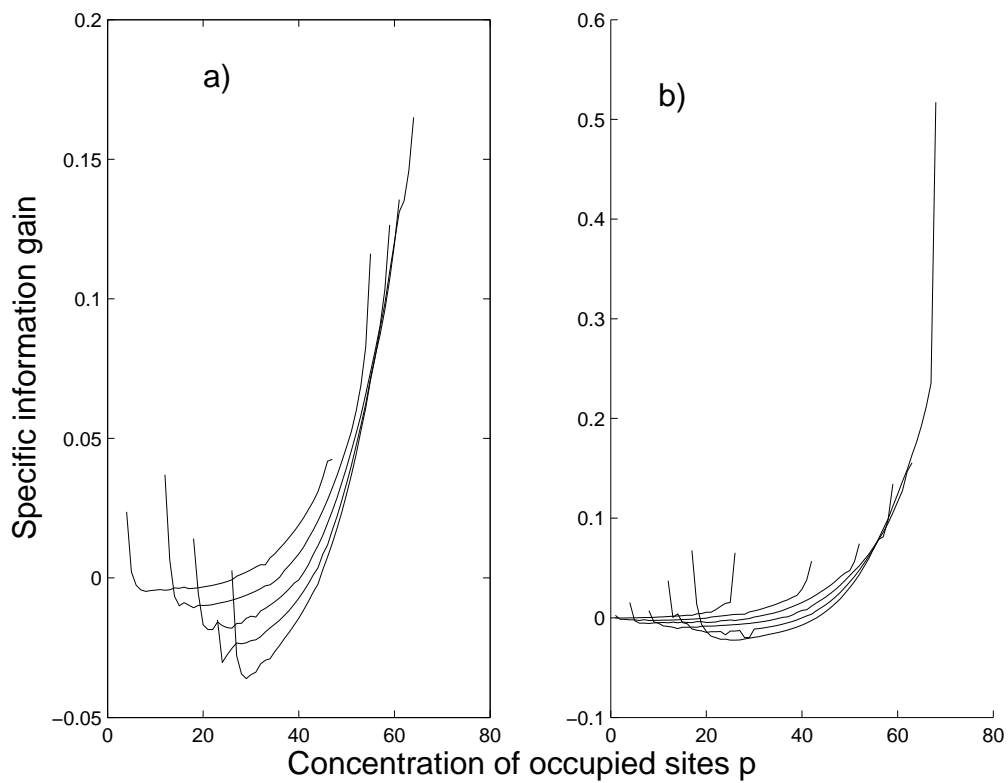}
\caption{\protect\label{Fig4} Panel a): Relative specific
information gain $I(p,p_{\xi})$ as a function of $p$ (in percent) for 
various values of $p_{\xi}= 2\%, 4\%, 6\%, 8\%, 10\%$ from left to right.
Panel b):  Relative specific information gain
$I(p,\xi)$ as a function of $p$ for 
various values of $\xi/L = 10\%, 20\%, 30\%, 40\%, 50\%, 60\%$ from left to
right.
}
\end{figure}

\clearpage

\begin{figure}
\includegraphics[width=14cm]{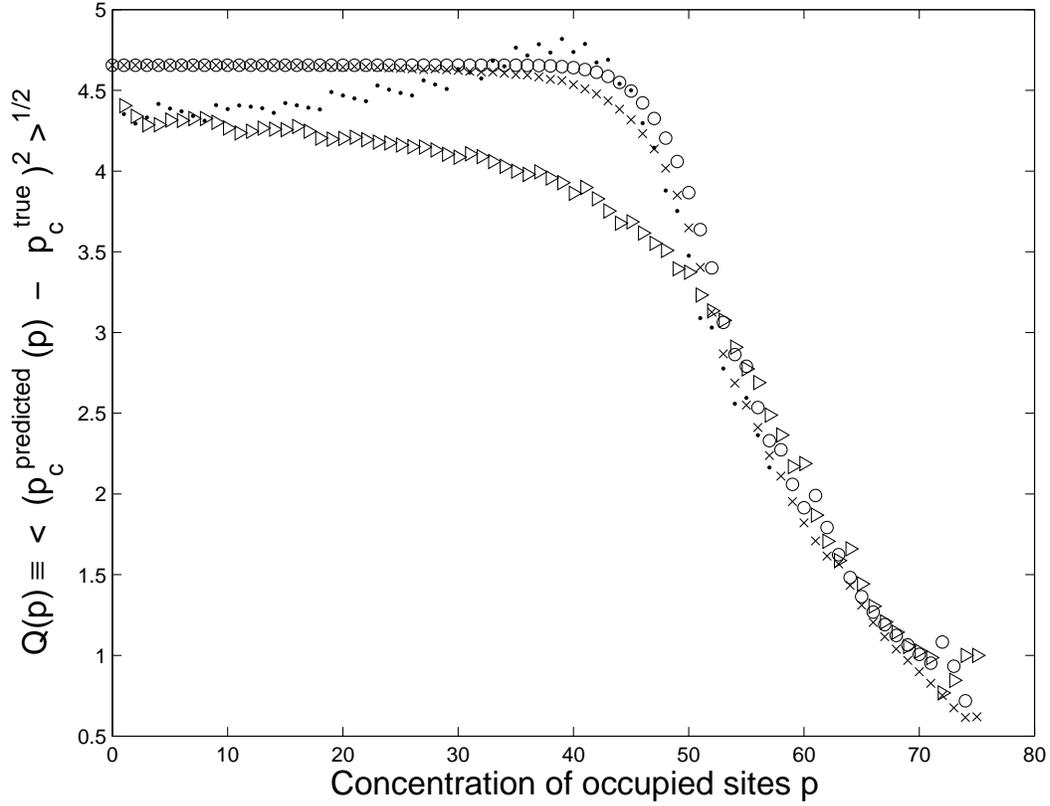}
\caption{\protect\label{Fig5b} 
RMS $Q(p, p_{\xi})$ (in percent) of the prediction
errors defined by (\ref{nvyhwy}) with (\ref{mgkkle;sa}),
for the quantiles $q=5\%$ and $q=95\%$ of the distribution of 
$\xi$ at fixed $p$, when using $P(p_c|p,\xi^q(p))$ as the predictor,
as a function of the damage parameter $p$ (in percent). Triangle: $q=5\%$; dots:
$q=95\%$; crosses: $q=50\%$; circles: $Q(p)$ obtained 
using the conditioning only on $p$. This RMS $Q(p, p_{\xi})$
should be compared with the standard deviation equal to $4.66\%$ of the 
unconditional distribution of percolation thresholds, to 
illustrate the gain in prediction accuracy deriving from
the added information.
}
\end{figure}

\clearpage

\begin{figure}
\includegraphics[width=14cm]{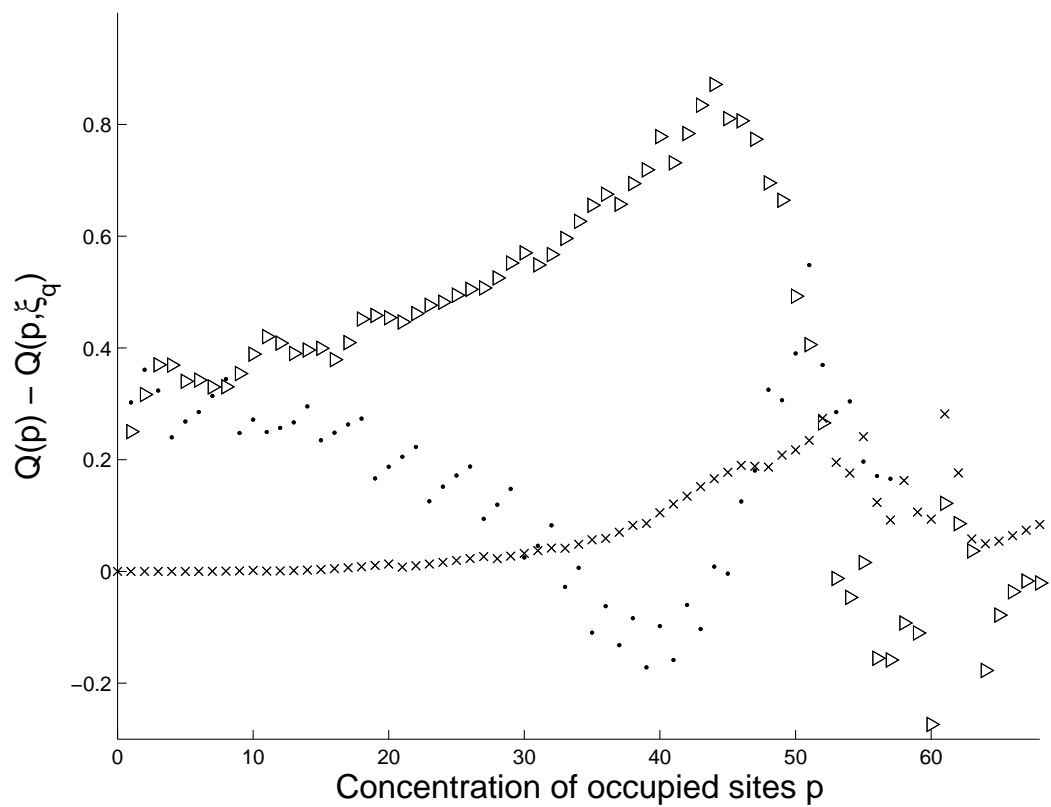}
\caption{\protect\label{Fig5ba} 
Gain in RMS $Q(p) - Q(p, \xi)$ when adding the information on $\xi$,
where $Q(p, \xi)$ is shown as the triangles ($q=5\%$), dots ($q=95\%$) and
crosses ($q=50\%$) and $Q(p)$ is shown in figure \ref{Fig5b} with the circles.
}
\end{figure}

\clearpage

\begin{figure}
\includegraphics[width=14cm]{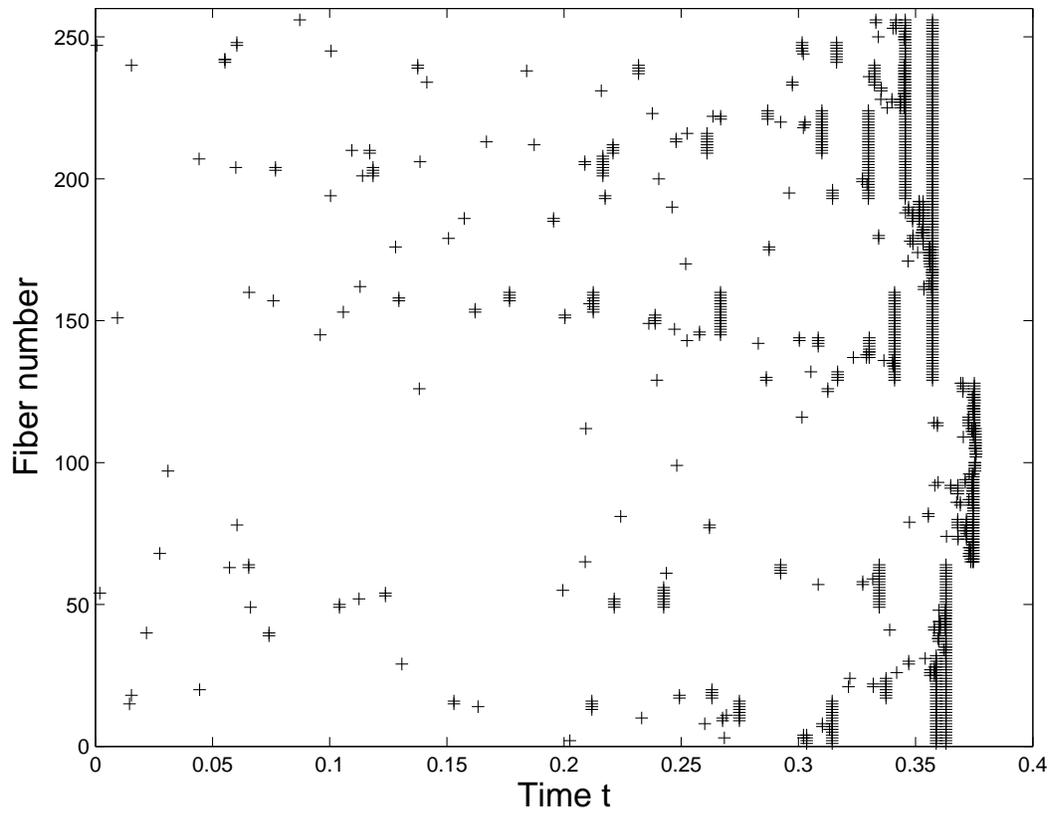}
\caption{\protect\label{fighierspacetime} A specific realization
of the space-time evolution of fiber damage for a system of
$2^8=256$ fibers. The fibers are numbered sequentially from $1$ to $256$
along the vertical axis. When a given fiber $i$ breaks at some time $t_i$,
a symbol $+$ represents the spatial position and failure time of this
event. 
}
\end{figure}

\clearpage

\begin{figure}
\includegraphics[width=8cm]{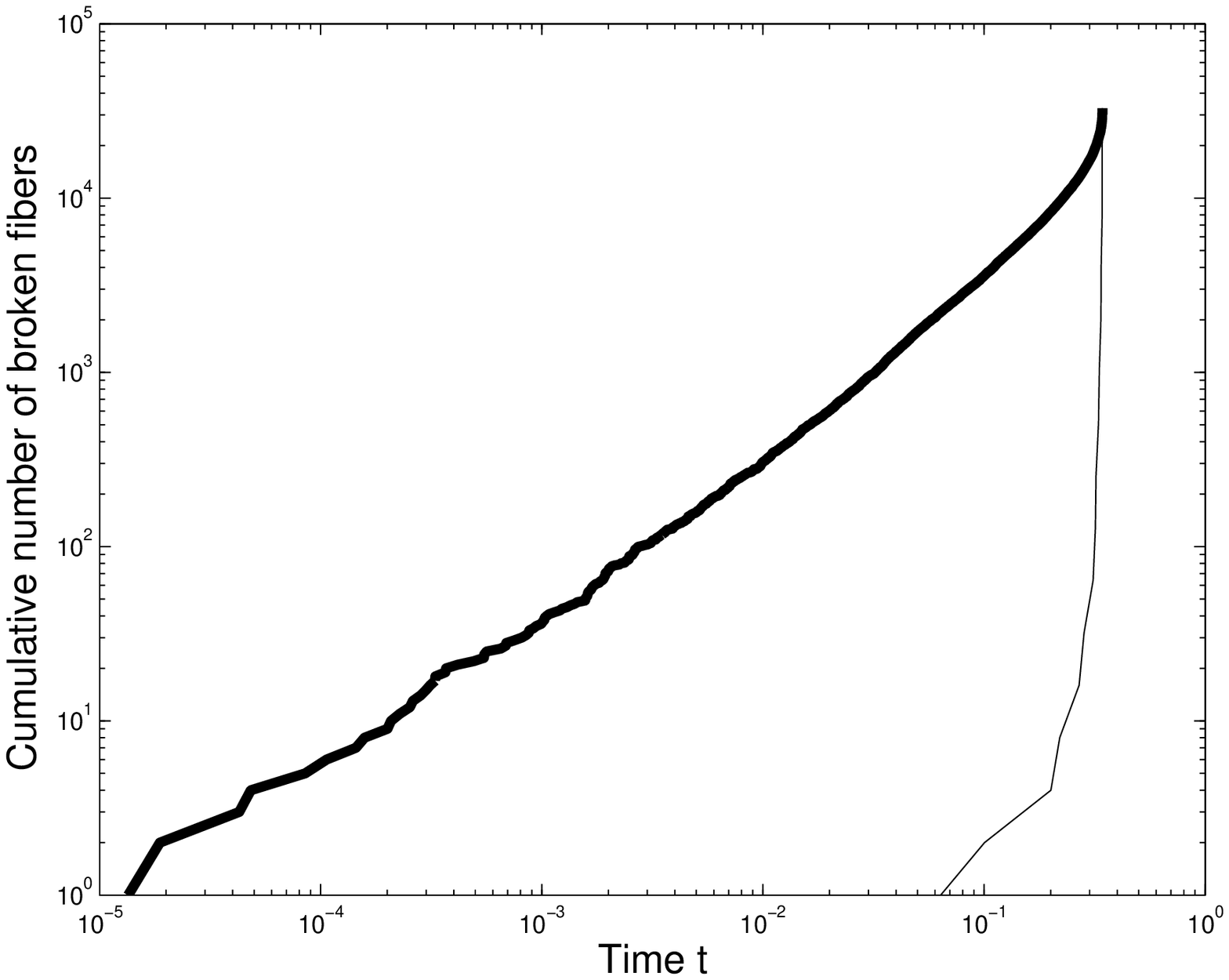}
\includegraphics[width=8cm]{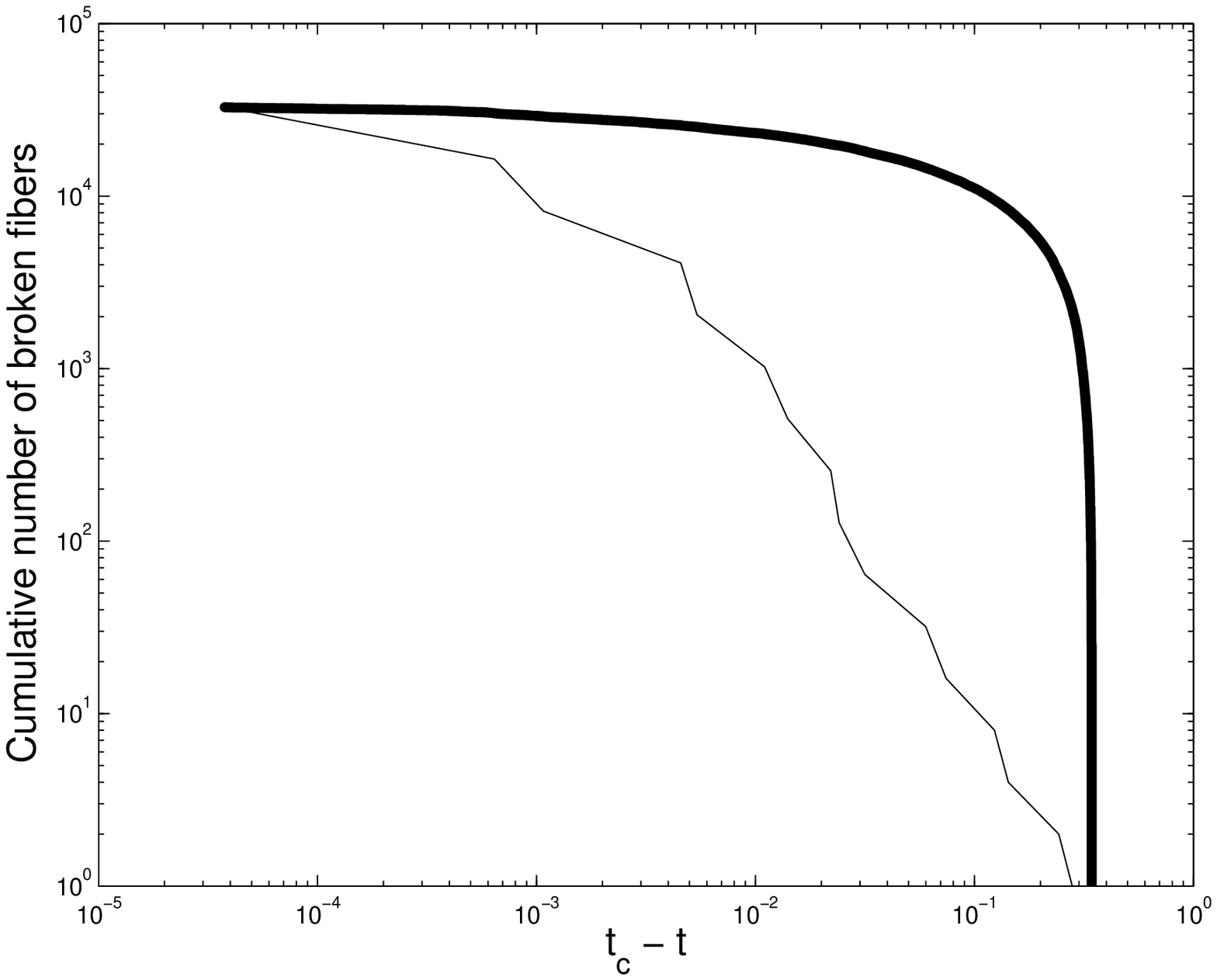}
\caption{\protect\label{cumcond} (Top) Two
different measures of the cumulative number of broken fibers as a
function of time $t$ for a given realization. The thick curve shows the
unconditional cumulative number of broken fibers. The thin curve shows
the (conditional) cumulative number of broken fibers, which
broke either within the largest crack identified up to time $t$ or within its complement in their
pair within the hierarchy. (Bottom) This graph shows the same two curves in log-log
scales with time $t$ replaced by $t_c-t$, where
$t_c$ is the global time of failure (only known at the end). This log-log representation
allows us to visualize the power law acceleration characterizing the final critical regime
before complete rupture, which is much more apparent in the conditional
cumulative number of broken fibers.
}
\end{figure}

\clearpage

\begin{figure}
\includegraphics[width=14cm]{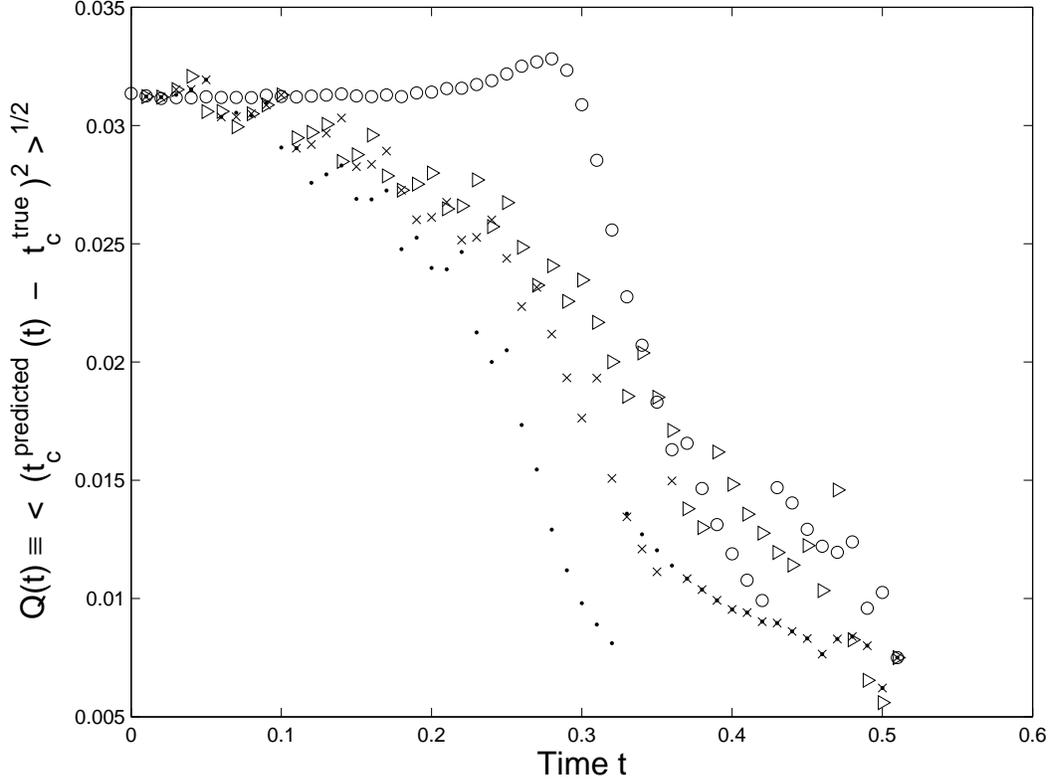}
\caption{\protect\label{Qtimehier} 
Root-mean-square (rms) $Q(t)$ of the error or difference between
predictions made at time $t$ of the global rupture time and the true
realized one as a function of time $t$ for $5$ distinct prediction
schemes using different conditioning, similarly to figure
\protect\ref{Fig5b} previously constructed for the percolation model.
The system used here has $2^8$ fibers and $\rho=2$.
The $o$ symbols correspond to a prediction at time $t$ of the failure time $t_c$
based solely on the information that the system has not yet broken.
For the other curves, we constructed the distribution of failure
times over $10^6$ realizations for the different conditioning.
The triangles correspond to the r.m.s. $Q(t)$ obtained by using the $5\%$
quantile of the distribution of failure times over these $10^6$
simulations. Specifically, for a given system, and at a given time $t$, 
we measure the size $\xi$ of the largest failed cluster and then read 
from the distribution of failure times for the same time $t$ and same cluster size $\xi$
the $5\%$ quantile that we take as the prediction for the failure time. 
Similarly for the $x$ and $.$  corresponding respectively
to the $50\%$ and $95\%$ quantiles. Note that in our system of 
$2^8$ fibers, there are $8$ possible sizes of ``cracks'' larger than $1$, namely
$2, 4, 8, ..., 128, 256$. These curves are obtained by averaging over
$10^5$ realizations. These RMS $Q(p)$ for the five prediction schemes
should be compared with the standard deviation equal to $0.0311$ of the 
unconditional distribution of failure times $t_c$, to 
illustrate the gain in prediction accuracy deriving from
the added information obtained from conditioning.
}
\end{figure}

\end{document}